\documentclass[epj]{svjour}
\def\tmx{$\mathrm{(TMTSF)_{2}ClO_{4(1-x)}ReO_{4x}}$}\,
\def\tmp{$\mathrm{(TMTSF)_{2}PF_{6}}$}\,
\def\tmc{$\mathrm{(TMTSF)_{2}ClO_{4}}$}\,
\,
\def\tmps{$\mathrm{(TMTTF)_{2}PF_{6}}$}\,
\def\tm{$\mathrm{(TMTSF)_{2}X}$}\,

\usepackage{graphicx}
\usepackage{amssymb}
\usepackage{epstopdf}
\begin{document}

\title{Field-induced confinement in  \tmc   \,under accurately aligned magnetic fields}
\author{N.Joo\inst{1,3} \and P. Auban-Senzier\inst{1} \and C.R. Pasquier\inst{1}\and  S. Yonezawa \inst{2}\and  R.  Higashinaka  \inst{2} Y.  Maeno \inst{2}  \and S. Haddad\inst{3} \and  S. Charfi-Kaddour\inst{3} \and M. H\'eritier\inst{1} \and K. Bechgaard\inst{4}\and D.J\'erome\inst{1}
}                     
\institute{Laboratoire de Physique des Solides (UMR 8502) - Universit\'e Paris-Sud, 91405, Orsay, France \and Department of Physics, Kyoto University, Kyoto 606-8502, Japan  \and  Facult\'e des Sciences de Tunis, LPMC, Campus Universitaire, 1060, Tunis, Tunisie \and Department of Chemistry, H. C. $\O$rsted Institute, Universitetsparken 5, DK 2100, Copenhagen, Denmark}
\date{Received: date / Revised version: date}
\abstract{We present transport measurements   along the least conducting (\textbf{c}) direction  of the organic superconductor \tmc \, performed under an accurately aligned magnetic field in the low temperature regime.
The experimental results reveal a two-dimensional confinement of the carriers in the (\textbf{a,b}) planes which is governed by  the  magnetic field component along the \textbf{b$^{\prime}$} direction. This 2-D confinement is accompanied by a metal-insulator transition for the \textbf{c} axis resistivity.
These data are supported by a quantum mechanical calculation of the transverse transport taking into account in self consistent treatment  the effect of the field on the interplane Green function and on the intraplane scattering time.
} 
\maketitle
\section{Introduction}
\label{intro}
\begin{figure}
\centering
\resizebox{0.75\columnwidth}{!}{%
\includegraphics{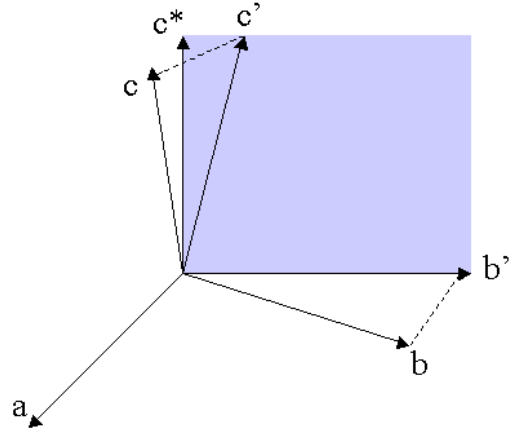}}
\caption{ Schematic representation of the different axes derived from the triclinic symmetry. }
\label{figure1.eps}
\end{figure}

The major unsolved problem for  quasi one dimensional (Q-1-D) organic conductors pertaining to the \tm\,  series is the determination of the mechanism leading to superconductivity emerging in almost all of them under or without  pressure. Based on the early finding of a great sensitivity of organic superconducting state to the presence of impurities and on the possibility of the critical fields being explained by an orbital limitation and consequently not destroyed by a paramagnetic mechanism, there have been a suggestion of triplet pairing superconductivity\cite{Oh04,Tomic83}. 
On the other hand  measurements of the critical fields performed along  the three principal axes \textbf{a}, \textbf{b$^{\prime}$} (normal to \textbf{a} in the (\textbf{a,b}) plane \textit{see} Fig. \ref{figure1.eps}) and \textbf{c$^{\star}$} down to 0.5K have revealed the Pauli limiting behavior and therefore the evidence of singlet pairing\cite{Murata87}.

Recent experiments of the dependence of the superconducting $T_c$ on the concentration of non magnetic defects in the solid solution  \tmx\,    have conclusively shown that the superconducting gap cannot keep the same sign over the whole Fermi surface\cite{Joo05,Joo04}. However, these experiments are still unable to provide a clue for the spin part of the superconducting pairing, namely  a singlet $d$-pairing or a triplet $p$ or $f$- pairing. Consequently, additional experiments sensitive to the spin part of the pairing are needed to move one step further. Such an experiment has been performed measuring the spin susceptibility in the superconducting state of \tmps \,via the $^{77}$Se Knight shift  \cite{Lee02}. The finding of a constant susceptibility within the error bars crossing the critical temperature has enabled the authors of the reference \cite{Lee02} to conclude in favour of an equal spin pairing, a form of the triplet pairing \cite{notetriplet}. Another approach has been used to probe the possibility of equal spin pairing \textit{via} the measurement of the critical fields of the superconducting state of \tmc.
Such data have been obtained recently using torque and resistive  determinations of the upper critical field of \tmc\, down to 25 mK under a precise alignment of the field along the \textbf{b$^{\prime}$} direction\cite{Oh04}. That work has shown that $H_{c2}^{b^\prime}$ can reach 5T at zero temperature  \textit{i.e} a factor about two above the Pauli limit  for singlet superconductivity. These data corroborate earlier works in the same material\cite{Lee95}  and also in the superconducting state of \tmp\, under pressure with the same field orientation\cite{Lee97}. However,  the \textbf{b$^{\prime}$} direction is somewhat peculiar in the \tm\, structure since it is a direction which is parallel to the (\textbf{a,b}) planes. Hence, as noticed by Lebed in 1986 a magnetic field-induced dimensional crossover is expected when the magnitude of the in-plane aligned field becomes of the order of the interlayer coupling so that the amplitude of the electron trajectories  becomes smaller than the interlayer distance\cite{Lebed86c}. For an in-plane alignment of the magnetic field $H_{c2}$ has been predicted to exhibit an upward curvature at low temperature with a possible reentrance of the superconducting phase in high fields for both singlet and triplet pairing although the latter is more evident\cite{Lebed86c,Dupuis93}. From the experimental point of view a negative temperature dependence of the resistance in the normal state has been observed for  $H$//\textbf{b$^{\prime}$} in both  \tmc\cite{Lee95} and \tmp\cite{Lee97}. The case of \tmp\, is disputed since the measurements require the stabilization of the superconducting phase under pressure and there  exists the possibility of a pressure regime in which superconductivity coexists with an insulating spin density wave state\cite{Vuletic}. In such a narrow pressure regime all components of the resistivity do reveal a negative temperature behaviour prior to the superconducting transition. However no such objection can be applied to \tmc\, in its very slowly cooled low temperature phase. Consequently, the insulating character of the resistivity which is observed in that compound at low temperature under field //\textbf{b$^{\prime}$} must be ascribed to a field-induced modification of the electron spectrum.
The present situation is such that a determination of the critical fields of \tmc \,  down to the lowest possible temperatures  with the field accurately oriented along the three respective crystal axes  on the same sample is still missing. 
\section{Experimental}
\label{}
\begin{figure}
\centering
\resizebox{0.75\columnwidth}{!}{%
\includegraphics{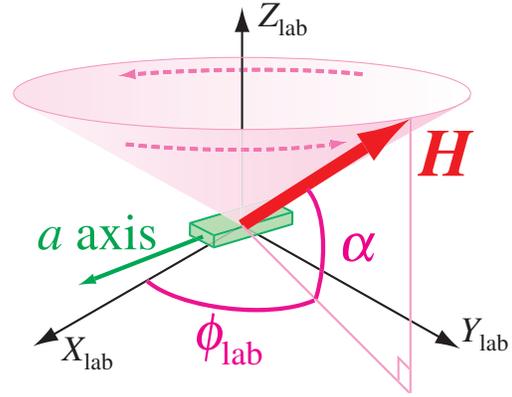}}
\caption{ Schematic representation of the sample orientation in the magnetic field. }
\label{figure2.eps}
\end{figure}

As a preliminary study, we have performed an experimental investigation of the role of the magnetic field along the principal axes in the (\textbf{a,b$^{\prime}$}) plane. Accurate alignment was enabled by a vector magnet system which provides high angular resolution \cite{Deguchi04} see Fig.  \ref{figure2.eps}. Furthermore,  we are also presenting a calculation explaining the electronic confinement in the (\textbf{a,b}) planes and  the non metallic behaviour of the \textbf{c} axis  resistivity observed at low temperature when a magnetic field is parallel to the \textbf{b$^{\prime}$} direction.

\begin{figure}
\centering
\resizebox{0.75\columnwidth}{!}{%
 \includegraphics{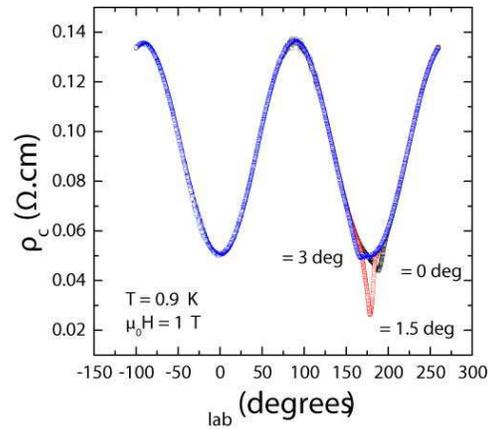}}
\caption{Procedure of field alignment : The two fold oscillation is dominated by the intrinsic in-plane anisotropy rather than the misalignment of $\pm$ 1.5 deg during the field rotation. The sharp dip at $\alpha$ =1.5 deg indicates accurate alignment of H//\textbf{a} axis. In this experiment,  field and temperature conditions have been purposedely chosen close enough to the critical superconducting conditions  in order to obtain a finite but non zero value for the resistance when the field is properly aligned.}
\label{figure3.eps}
\end{figure}

The first step consists in the determination of the principal axes from the magnetoresistance using the anisotropic properties of superconducting state. The sample has been glued on a thin sapphire plate for a good thermal contact with the reservoir. Keeping the sample fixed,   the field of 1 Tesla (at the temperature of 0.9K) has been rotated around the vertical axis in a plane making an angle $\alpha$ with respect to the plane of the sample. The field and the temperature have been chosen following the early data of reference\cite{Murata87} in order to keep the sample superconducting and normal for  alignments along \textbf{a} and \textbf{b$^\prime$} respectively.  The angle $\alpha$ providing the accurate alignment of the field parallel to the (\textbf{a,b}) plane is the one which reveals the sharpest 
and deepest  drop of the resistance in the rotation pattern for a given angle $\phi$  which marks therefore the position of the \textbf{a} axis, see Fig. \ref{figure3.eps}. 


\begin{figure}
\centering
\resizebox{0.85\columnwidth}{!}{%
 \includegraphics{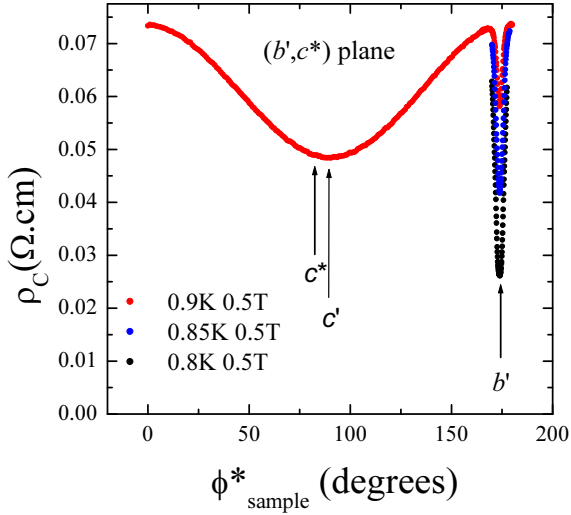}}
\caption{Field rotation in the plane normal to the sample \textbf{a} axis. The \textbf{a} axis direction was determined by the procedure in Fig.\ref{figure2.eps}. The sharp dip is associated with the onset of superconductivity with the field H//\textbf{b$^{\prime}$}. It also indicates that a resistivity minimum occurs for H//\textbf{c$^{\prime}$} which is 7 deg away from the \textbf{c$^{\star}$} direction. The non zero value of the resistance for the \textbf{b$^{\prime}$} direction is due to the reasons explained in the caption of Fig. \ref{figure3.eps}.}
\label{figure4.eps}
\end{figure}
\begin{figure}
\centering
\resizebox{0.75\columnwidth}{!}{%
  \includegraphics{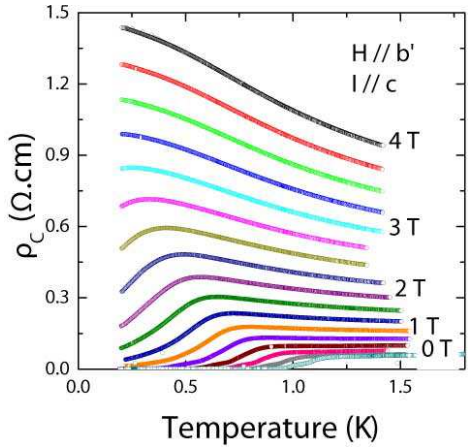}}
  \caption{ Transverse resistivity versus temperature for different values of the magnetic field H//\textbf{b$^{\prime}$} axis from 0 to 4T.}
\label{figure5.eps}
\end{figure}
\begin{figure}
\centering
\resizebox{0.75\columnwidth}{!}{%
 \includegraphics{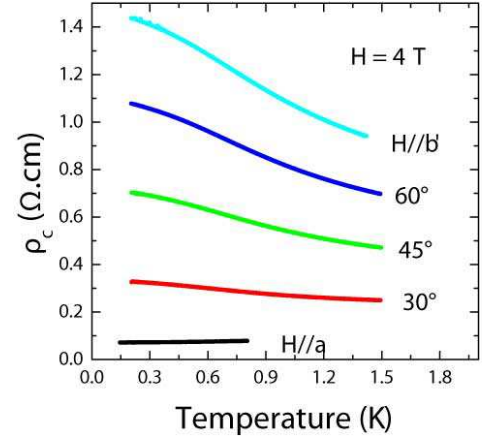}}
\caption{ Temperature dependence of the magnetoresistivity for different angles between \textbf{a} and \textbf{b$^{\prime}$} axes under 4 T. }
\label{figure6.eps}
\end{figure}

Next, a rotation of a 0.5 T field has been achieved in a plane perpendicular to the \textbf{a} direction previously determined at different temperatures between 0.8 and 0.9 K. According to Fig. \ref{figure4.eps} a very sharp  and strongly temperature dependent resistance minimum is observed at an angle corresponding to the  direction of the \textbf{b$^{'}$} axis  while the position of the shallow minimum reveals the direction of the \textbf{ c$^{\prime}$} axis  (the projection of \textbf{c} on the plane normal to \textbf{a}) and not of the\textbf{ c$^{\star}$} axis. 
In a second step the temperature dependence of the resistance has been measured along the three previously determined axes at various fields up to 4 T for \textbf{a} and \textbf{b$^{\prime}$} and up to 0.18 T for \textbf{c$^{\star}$} . As far as the \textbf{a} and \textbf{c$^{\star}$} directions are concerned the resistance always exhibits a transition  between a superconducting state at low temperature and a metallic state at high temperature. Such a behaviour is no longer followed for the \textbf{b$^{'}$} axis, Fig. \ref{figure5.eps}, as a very strong magnetoresistance is observed at high temperature with the resistance changing from a metallic to a non metallic behaviour at fields above $\approx$  1T.  In order to better characterize the phenomenon of localization observed on the \textbf{c}-axis transport under field several temperature sweeps have been recorded at various angles of the field in the (\textbf{a},\textbf{b$^{\prime}$}) plane at 4T, \textit{see} Fig.\ref{figure6.eps}. 

\begin{figure}
\centering
\resizebox{0.75\columnwidth}{!}{%
\includegraphics{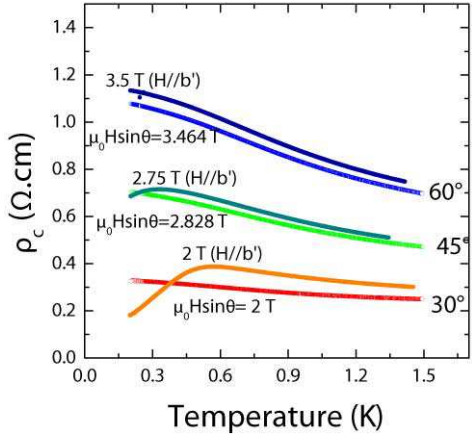}}
\caption{ Contribution to the transverse localisation of the component of the magnetic field along the \textbf{b$^{\prime}$}-axis and comparison with similar values  when H//\textbf{b$^{\prime}$}. }
\label{figure7.eps}
\end{figure}

Figure.\ref{figure6.eps} clearly shows that the metallic behaviour under the magnetic field of 4T is only observed when the field is aligned along \textbf{a}. Actually, comparing the data in Figs. \ref{figure5.eps} and \ref{figure6.eps} we can show that the localization is  governed by the component of the field along the \textbf{b$^{\prime}$} direction, see Fig. \ref{figure7.eps}. Furthermore,  it is interesting to see that  the  longitudinal resistivity $\rho_a$ retains its metallic character down to low temperatures even at high fields up to 25T  parallel to the \textbf{b$^{\prime}$} axis, see the insert in Fig.1 of reference \cite{Behnia95} . Therefore, the electronic confinement within the (\textbf{a},\textbf{b$^{\prime}$}) planes is accompanied by a localisation only for the conduction along the \textbf{c}-axis.

The coupling along the least conducting direction is not known very accurately but it is of the order of $\approx 5K$ or less. Hence, it is likely that the band model may breakdown in the  \textbf{c$^\star$}  direction due to thermal fluctuations when k$_{B}T$ exceeds $t_c$,namely around liquid helium temperature. This situation is supported by the finding of a small Drude component in the low frequency optical conduction only at low temperature\cite{Henderson99} and also by a resisitivity which is smaller than the maximum metallic resistivity ( 0.06 $\Omega cm$)\cite{Mott71}  only when the temperature is in the liquid helium range\cite{Joo05}. 
 
 In a preliminary approach one can view the \textbf{ c$^\star$}-axis transport as diffusive with a  probability for interlayer hopping  given by $1/\tau_{\bot} \approx t_{\bot}^{2} \tau_{a-b}$ where $ \tau_{a-b}$ is an average lifetime in the (\textbf{a,b}) planes, extending a model of interchain to interplane hopping\cite{Weger,Jerome82}. The \textbf{c$^{\star}$} axis transport thus reads $\sigma_{c^\star} \approx t_{\bot}^{2} \tau_{a-b}$. Assuming an interplane coupling insensitive to the magnitude of the magnetic field along \textbf{b$^{\prime}$} the magnetoresistance along \textbf{c$^\star$} should in turn follow that of $ \tau_{a-b}$ (or $\rho_a$). However,  at the temperature of 1.2K for instance, the magnetoresistance for $\rho_{ c^\star}$ amounts to a factor 17.5 in a field of 4T  which is \textit{at variance} with the factor 3.5 observed for the \textbf{a} axis component at similar values of temperature and field \cite{Kang}. Consequently, we can infer that the interlayer coupling must be affected by the magnetic field and possibly drops by a factor  5 under a field of 4T along \textbf{b$^{\prime}$}  as a result of a field-induced confinement and in turn becomes two dimensional (2D) under field.

\section{The confinement scenario: theory}

In the following, we propose a simple interpretation of the experiments
reported above and those of Danner \textit{et al.}\cite{Danner97}. In
particular, we discuss the effect of a magnetic field applied parallel to
the (\textbf{a,b}) plane, which does not modify the metallic conductivity in
the \textbf{a} direction, but induces a metal insulator crossover along the 
\textbf{c} direction. Let us precise that we do not discuss the possibility
of a superconducting order, and we discard it. We are, here, only interested
in the metallic or insulating character of the phase. Our model relies on 
\textit{three essential features, }which determine the physics of the
system, independantly of the details of the model : \textit{(i) }the
electrons can be described in a \textit{Fermi liquid picture. }Given the
experimental conditions, \\  \textit{i. e.} temperature much smaller than the
interplane electron transfer integral $t_{c}$, the electronic regime is
safely assumed to be three-dimensional and the Fermi liquid approach to be
valid ; \textit{(ii) }the strong anisotropy of the crystal structure is
reflected in the\textit{\ strong anisotropy of the electronic properties.\ }
The transfer integrals in the three directions exhibit a strong hierarchy,
leading to an \textit{open Fermi surface} both in the $\left(
k_{a},k_{b}\right) $ plane and in the $\left( k_{a},k_{c}\right) $ plane ; 
\textit{(iii)} we propose that the effects reported here are due to \textit{
an orbital effect of the field}, since only the \textbf{b} component of
the field is relevant\textit{\ . }As will be discussed below, both in a
semi-classical approach and in a fully quantum mechanical treatment, such an
orbital effect tends to reduce the dimensionality of the electron system,
therefore explaining the transport properties reported here.

To discuss this effect, we consider a simple model described by a mean field
single particle Hamiltonian, with a very small interplane electron transfer
and a tight binding dispersion relation : \newline
\begin{eqnarray*}
\epsilon \left( \vec{k}\right)  &=&v_{F}\left( |k_{x}|-k_{F}\right)
-2t_{b}\cos k_{y}b-2t_{b}^{\prime }\cos 2k_{y}b \\
&&-2t_{c}\cos k_{z}c,
\end{eqnarray*}
where we have taken a linear dispersion around the Fermi level along the
chains direction \textbf{a} corresponding to the highest conductivity and
with a Fermi velocity $v_{F}$. The dispersion in the transverse directions (
\textbf{b}, \textbf{c}) is described within a tight binding picture where $
t_{b}$ and $t_{b}^{\prime }$ are respectively the hopping integrals to the
first and to the second nearest neighbors in the \textbf{b} direction
whereas $t_{c}$ describes the hopping along the least conducting axis 
\textbf{c} perpendicular to the conducting plane (\textbf{a},\textbf{b}).
\newline
For the sake of simplicity, we have considered here an orthorhombic crystal
structure, although the real one is triclinic, but such a simplification
should not modify qualitatively the physical picture. The highly anisotropic
bandwidths given by: $t_{a}:t_{b}:t_{c}\sim 3000$ K : 200 K: 10 K, where $
t_{a}$ is the hopping integral in the \textbf{a} direction, lead to a
quasi-1D character to the (TMTSF)$_{2}$X compounds.\ 

At ambient pressure and above the superconducting transition temperature $
T_{c}\sim $ 1 K, the ground state of the (TMTSF)$_{2}$ClO$_{4}$ is metallic.
We consider a parallel magnetic field $\vec{H}\left( 0,H,0\right) .\;$Let us
first discuss a \textit{semiclassical approach, }which\textit{\ } gives an
intuitive understanding of the phenomenon. The orbital effect of a field
parallel to the \textbf{b} direction will reduce the $t_{c}$ hopping
process, leading to a 3D-2D crossover for an applied field of the order of $
t_{c}.$ This dimensional crossover can be understood by considering the
semi-classical equation of motion of the electron wave packet for H//\textbf{
b} on the Fermi surface: $d\vec{k}/dt=e(d\vec{r}/dt)\wedge H/\hbar c_{0}$.
The wave packet trajectory in $k$ space is linear along $k_{c},$ with
transverse periodic oscillations with an amplitude $\delta k_{x}=$  $
4t_{c}/v_{F}.$This gives rise, in real space, to a linear trajectory along
the \textbf{a} direction modulated by periodic oscillations along the 
\textbf{c} axis with an amplitude $\delta z=$ $4ct_{c}/ev_{F}H$ which
decreases as the magnetic field increases, making the electrons confined in
the (\textbf{a,b}) plane. For field strength of the order of $t_{c},$ the
electrons get confined in a single plane perpendicular to $c.$ In this
semi-classical picture, we do not expect a strong effect of the field on the
conductivity along the \textbf{a} direction. On the contrary, the electron
confinement in the (\textbf{a,b}) planes should correspond to a
metal-insulator crossover in the \textbf{c} direction. Such a \textit{field
induced confinement}, which should not be confused with a \textit{field
induced localization}, allows to understand the paradox of a qualitatively
different field effect on the conductivity along the \textbf{a} and \textbf{c
} directions. As a result the behavior of the \textbf{c} axis resistivity is
expected to change from metallic to non metallic as the temperature
decreases.


The dimensional crossover has also been described by Strong \textit{et al.} 
\cite{Strong94} as the consequence of an interlayer decoupling which happens
above a critical field separating a 3D Fermi liquid, where interlayer
hopping is coherent, from a 2D non Fermi liquid where the hopping is
incoherent. 

Such a model is somewhat similar to another one discussed earlier \cite
{Gorkov84,Heritier84} where the field is perpendicular to the most
conducting planes. For such a geometry, it has been shown that the orbital
effect of the field induces a one-dimensionalization of the electron motion,
because it cancels the effect of $t_{b}^{\prime }$ and because the transfer
integral in the third direction $t_{c}$ can be neglected above a threshold
field of a few Teslas. This model has successfully explained the overall
features of the well known FISDW phases. However, the physics for a field
orientation perpendicular to the (\textbf{a,b}) plane is different from that
of parallel field H//\textbf{b}. First, in the latter geometry, the transfer
integral along the field direction is not $t_{c}$ as in the former case but $
t_{b}$ which is much larger than the field, thus precluding the possibility
of a SDW instability.

The semi-classical argument given above is fully confirmed by a quantum
mechanical treatment.\ We consider a magnetic field described in the gauge $
\vec{A}\left( 0,0,-Hx\right) $ and we make the corresponding Peierls
substitution $\vec{p}\rightarrow \vec{p}-\frac{e}{c_{0}}\vec{A}$, ($c_{0}$
is the light velocity). Because of the $x$ dependence of $A,$ it is more
convenient to use a mixed $\left( x,k_{y}\right) $ representation\cite
{Gorkov84} .Our purpose is a theoretical study of the conductivity along the 
\textbf{c} direction. The latter mainly depends on two different parameters
: the first is the electron transfer along the \textbf{c} direction, which
we can study by calculating the interplane Green's function ; the second is
the in-plane electron scattering time. To discuss correctly the conductivity
in the \textbf{c} direction, it is essential to take into account the fact
that both parameters are strongly affected by the orbital effect of the
field. 

Let us first calculate the the\textit{\ independant electron Green function} 
$G_{++}$ near the right-hand Fermi surface sheets,the equation of motion of
which is given by,
\begin{eqnarray}
\large\left\{\right.i\omega_n&&+iv_F\frac{d}{dx}-2t_b\cos k_yb -
2t_b^{\prime}\cos2k_y b\nonumber\\
&&\left.-2t_c\cos \left(k_z+\frac{eHx}{c_{0}}\right)c\right\}\nonumber\\
&&g_{++}\left(i\omega_n,k_y,k_z,x,x^{\prime}\right)=\delta(x-x^{\prime}),
\label{g}
\end{eqnarray}
where $g_{++}\left(i\omega_n,k_y,k_z,x,x^{\prime}\right)= \\ e^{-ik_F(x-x^{\prime})}
G_{++}\left(i\omega_n,k_y,k_z,x,x^{\prime}\right)$.

The integration of Eq.\ref{g} is straightforward.  We obtain for $\omega
_{n}(x-x^{\prime })>0$: 
\begin{eqnarray*}
&&G_{++}\left\{ i\omega _{n},k_{y},k_{z},x,x^{\prime }\right) =\frac{\mathrm{
sign}\;\omega _{n}}{iv_{F}}\exp i\left\{ {}\right. \frac{i\omega
_{n}(x-x^{\prime })}{v_{F}} \\
&&\left. +k_{F}(x-x^{\prime })-\frac{1}{v_{F}}\left[ 2t_{b}\cos
k_{y}b+2t_{b}^{\prime }\cos 2k_{y}b\right] (x-x^{\prime })\right\}  \\
&&-\frac{4t_{c}}{\omega _{c}}\cos \left( k_{z}c+\frac{G_{c}(x+x^{\prime })}{2
}\right) \sin \left( \frac{G_{c}(x-x^{\prime })}{2}\right) \left. {}\right\}
,
\end{eqnarray*}
where $G_{c}=eHc/c_{0}$ is the magnetic wave vector and $\omega
_{c}=v_{F}G_{c}$ is the magnetic energy.\ 

To bring out the $z$ dependence of the Green function, we have taken the
Fourier transform of \newline
$G_{++}\left( i\omega _{n},k_{y},k_{z},x,x^{\prime }\right) $ with respect
to $k_{z}$. We substitute $i\omega _{n}$ by $\omega +i\eta $, where the
imaginary part $\eta $ is related to the quasi-particle lifetime $\tau $ by $
\eta =\hbar /\tau $. we obtain: 
\begin{eqnarray}
&&G_{++}\left( \omega ,k_{y},z,x,x^{\prime }\right) =\frac{1}{2\pi }\frac{
\mathrm{sign}\;\omega }{iv_{F}}\exp \Big\{\frac{-\eta (x-x^{\prime })}{v_{F}}
\Big\}  \nonumber \\
&&\exp i\Big\{ \frac{\omega (x-x^{\prime })}{v_{F}}+k_{F}(x-x^{\prime })-
\frac{1}{v_{F}}\big[2t_{b}\cos k_{y}b  \nonumber \\
&&+2t_{b}^{\prime }\cos 2k_{y}b\big](x-x^{\prime })\Big\} \times \int_{-
\frac{\pi }{c}}^{\frac{\pi }{c}}\exp i\Big\{k_{z}z-\frac{4t_{c}}{\omega _{c}}
\times   \nonumber \\
&&\cos \left( k_{z}c+\frac{G_{c}(x+x^{\prime })}{2}\right) \sin \left( \frac{
G_{c}(x-x^{\prime })}{2}\right) \Big\}dk_{z},  \label{eta}
\end{eqnarray}

\begin{figure}[tbp]
\centering
\resizebox{0.95\columnwidth}{!}{\includegraphics{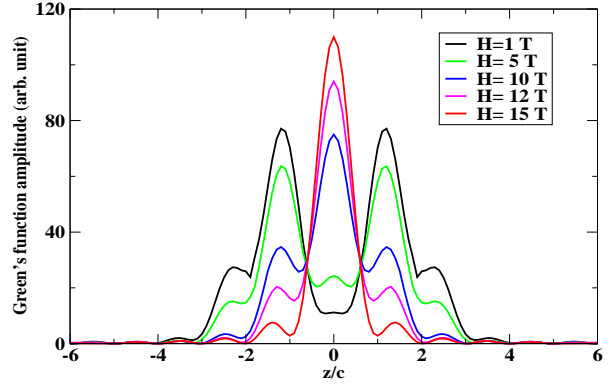}}
\caption{{\protect\small $z$ dependence of the amplitude of the Green
function for different values of the magnetic field.}}
\label{Green3H.eps}
\end{figure}

In the following we give the numerical results obtained from the method
explained here. In Fig.\ref{Green3H.eps} we have depicted the amplitude of $
G_{++}\left( \omega ,k_{y},z,x,x^{\prime }\right) $ as a function of $z$. As
the magnetic field increases, the electron Green's function gets more and
more confined by the orbital effect. At a field of 1 Tesla, its amplitude is
restricted in a volume between the planes z = - 2 and $z$ = + 2, which is,
already, a strong confinement. However, because of the strongly oscillating
character of the Bessel functions entering the expression of the Green's
function, an interference effect occurs at this moderate field, leading to
strong peaks at $z$ = +2 and $z$ = -2, but to a decrease of the amplitude in
between, around $z$= 0, compared to that at $z$ = 2. For larger fields,
namely of the order of 10 Tesla, this interference effect disappears : the
amplitude peak is centered at $z$ = 0, as expected and, at 15 Tesla, the
Green's function is almost entirely confined in the plane $z$ = 0. 

However, the electrical resistivity also depends on a second factor, the 
\textit{in-plane electron relaxation time }$\tau ,$ which is, of course,
related to the imaginary part of the frequency $\eta =\hbar $ $/$ $\tau $ 
It is worth stressing, that in Eq.\ref{eta}, we have taken into account the
dependence of the $\eta $ factor on both temperature and magnetic field.
This dependence is derived using the well known temperature dependence of a
Fermi liquid resistivity at low $T$, which is governed by the
electron-electron interactions. In the case of an isotropic Fermi liquid,
the quasi-particle lifetime $\tau $ obeys  the law: 
\[
\frac{1}{\tau }\propto T^{2}n(E_{F}).
\]
Here $n(E_{F})$ is the density of states at the Fermi level. For layered
conductors, the Fermi surface is anisotropic and the in-plane density of
states is $k_{z}$ dependent for a fixed $k_{x}$ ($k_{x}=k_{F}$). The $\eta $
factor is then given by: 
\[
\eta (k_{z})=-2\mathrm{Im}G_{++}\left( \omega
=E_{F},k_{x}=k_{F},k_{z}\right) T^{2}.
\]
\newline
Therefore the field effects on the electron transfer along the \textbf{c}
direction and on the in-plane electron relaxation time\textit{\ }$\tau $ are
intimately coupled together. First, the field, by confining the electrons in
the planes perpendicular to \textbf{c}, strongly reduces the amplitude of
the electron transfer from plane to plane in the \textbf{c} direction. But,
at the same time, the field, by inducing a strong electron confinement,
localizes the electron density in a narrower distance along \textbf{c}.
Straightforwardly, such a strong increase of the electron density in the
planes perpendicular to \textbf{c} induces a subsequent increase of the
electron scattering rate $\hbar /\tau $. These two effects should,
therefore, be discussed together by a self consistent procedure, which, as
far as we know, has not been addressed so far. Such a self consistency is
essential for the soundness of the method, since the electron confinement as
well as the increase of the scattering rate are induced by the same orbital
effect of the field. The electron scattering rate, indeed, strongly depends
on the in-plane electron density. The latter is calculated by the electron
Green's function, in which the imaginary part of the frequency argument is
precisely the electron scattering rate. As we shall see in the following,
such calculations describe successfully the experimental data discussed in
this paper and in ref.\cite{Danner97}. We have indeed calculated the
transverse component $\sigma _{zz}$ of the electronic conductivity given by
the Kubo formula, in which the scattering rate $\eta $ plays an essential
role \cite{Grigoriev03,Mahan90},

\begin{eqnarray*}
\sigma _{zz} &=&\frac{e^{2}\hbar }{\Omega }\int dk_{y}\int
dk_{z}v_{z}^{2}\int dE \\
&&\times \int \frac{d\epsilon }{2\pi }\left[ 2\mathrm{Im}
G_{R}(k_{y},k_{z},E,\epsilon )\right] ^{2}\times \left[ -n_{F}^{\prime
}(\epsilon )\right] 
\end{eqnarray*}
where $E=v_{F}/(x-x^{\prime })$, $\Omega $ is a normalization factor, $v_{z}=
\partial \epsilon (k_{y},k_{z},\omega _{c})/\hbar \partial k_{z}$ is the
electron velocity in the $z$ direction and $n_{F}^{\prime }(\epsilon )$ is
the derivative of the Fermi distribution function. $G_{R}(k_{y},k_{z},E,
\epsilon )$ is the retarded Green function. 

\begin{figure}
\centering
\resizebox{0.85\columnwidth}{!}{
\includegraphics{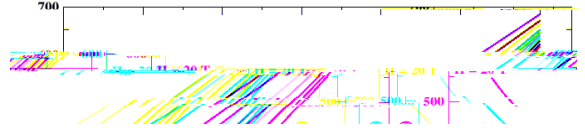}}
\caption{ Resistivity versus temperature for different values of the
magnetic field.}
\label{Rzz.eps}
\end{figure}

In Figure.\ref{Rzz.eps} we plot the
temperature dependence of the resistivity $\rho _{c}=1/\sigma _{zz}$ for
different values of the magnetic field. In the low temperature range, $\rho
_{c}$ changes from the metallic regime for $H$ below a threshold value $
H_{co}\sim 1T$ to an insulator state for higher fields. Hence, the
resistivity exhibits a minimum in temperature. The larger the magnetic
field, the larger the temperature at which the resistivity is minimal. These
features are quantitatively consistent with the experimental data of Fig.\ref
{Rzz.eps} and with the results of Ref\cite{Danner97}.
However, it should be stressed that for $\omega _{c}<T$ ( \textit{i. e.} above 30 K
for fields of the order of 10 T), the effective dimensionality of the system
is 2D. In this regime, the third direction (governed by $t_{c}$) is erased
by thermal fluctuations. In this limit, the effect of the field on the
conductivity should vanish. Moreover, the in-plane quasiparticle life time
should not depend anymore on the magnetic field, since the electron motion,
in this temperature range, is actually confined in a single (\textbf{a,b})
plane. 

This confinement scenario cannot  
explain the tendency towards saturation for the resistance measured at low temperature under high fields in Fig.\ref{figure5.eps} which is possibly due to  incipient superconducting fluctuations effects or impurities in a narrow gap insulator.

It is worth stressing that we did not address in our model the effect of the
superconducting fluctuations which are enhanced as the temperature decreases
towards the superconducting transition temperature $T_c\sim $ 1.2 K. This
study goes beyond the scope of this work.

\section{Conclusion}
A quantum calculation of the conduction along the \textbf{c} axis of \tmc has shown that the insulating character of the resistance which is observed at low temperature under a magnetic field larger than $\approx 1T$ aligned along the \textbf{b$^{\prime}$} direction   can be explained by a confinement of the carriers within the (\textbf{a,b}) planes\cite{noteLebed}. Consequently, it is now clear that the resistive signature attributed to the onset of  superconductivity  with a diverging critical field (H//\textbf{ b$^{\prime}$}) below 0.2K \cite{Oh04} must be reexamined in the context of superconductivity arising in a metallic multilayered compound. Long range ordered superconductivity might then be replaced by strong fluctuations  at very low temperature.
\section{Acknowledgments}

N.Joo acknowledges the french-tunisian cooperation CMCU (project 04 G1307 ) and a support from the University of Kyoto for her stay in Japan.

%
%
%

\begin{thebibliography}{10}
\bibitem{noteLebed}
During the completion of this work, we have been aware of a similar field
  induced 3D-2D crossover, discussed independently by A.G. Lebed in a recent
  paper \cite{Lebed05}. The author calculated the wave function, using a
  different gauge and introducing a $z$ dependent potential energy but did not
  discuss the transport measurements in the low field limit reported here. The
  author suggested to make use of the localization effect, in particular the
  high field limit ($H=$30-45 T) to investigate the Fermi surface and the
  validity of the Fermi liquid picture in layered organic and high-$T_c$
  materials.
\end{thebibliography}

\end{document}